\documentclass[12pt, nofootinbib]{revtex4}
\usepackage{amsmath,amssymb}
\usepackage{verbatim,graphicx}

\newcommand\ket[1]{| #1\rangle}
\newcommand\bra[1]{\langle #1|}
\newcommand\BR{\mathbb{R}}
\newcommand\BZ{\mathbb{Z}}
\newcommand{\BC}{\mathbb {C}}

\def\Tr{\textrm{Tr}}

\newcommand{\beq}{\begin{equation}}
\newcommand{\beqs}{\begin{equation*}}
\newcommand{\eeq}{\end{equation}}
\newcommand{\eeqs}{\end{equation*}}
\newcommand{\CO}{  {\cal O}  }

\begin{document}
\setlength{\unitlength}{1mm}
\title{Giant gravitons: a collective coordinate approach}

\author{David Berenstein }
\affiliation { Department of Physics, University of California at Santa Barbara, CA 93106}

\begin{abstract} In this paper I describe a collective coordinate approach to the study of giant graviton states and their excitations in various field theories. The 
method simplifies considerably the understanding of emergent gauge symmetry of these configurations, as well as the calculation of the spectrum of strings stretched between 
the giant gravitons. There is a limit where these results reproduce the one loop dispersion relation for giant magnons.
I also show that this method gives rise to a simple geometric interpretation of a Higgs mechanism for the emergent gauge symmetry which parallels the holographic dual realization of these sates: the effective  Higgs condensate is the geometric separation of D-branes in the collective coordinate geometry. 
\end{abstract}

\maketitle

\section{Introduction }
\label{S:Introduction}

The AdS/CFT correspondence \cite{Malda} is a remarkable duality between ordinary field theories and quantum gravity in higher dimensions. The simplest such quantum gravity theories are usually a string theory, as originally proposed by 't Hooft \cite{'tHooft} where the string theory is obtained by resuming planar diagrams.
The gravity theory can sometimes be semiclassical in the strong 't Hooft coupling limit. 
 Since the gravity theory lives in higher dimensions, one can think of quantum gravity and geometry in higher dimensions as an emergent collective dynamics arising from strong interactions in field theory. As such, the duality promises to resolve long-standing questions in the formulation of quantum gravity and it also promises to give a new understanding of how geometry might break down in high energy collisions. 
 
 String theories are not just theories of strings. There are a host of non-perturbative defects that string theories have. The simplest such defects are D-branes \cite{DLP,Pol}. For example there are cases where a gauge theory is Higgsed, and this is realized as a stack of parallel branes, with one extra brane separated by a finite distance to the rest.  The Higgs branch physics affects the dictionary relating the gravity fields to those of the field theory \cite{KW}. One can also add infinite probe branes to a given configuration, giving rise to a field theory with defects, or with flavors (see \cite{Kruczenski:2003uq} for example). 
 
 D-branes can also be realized by finite energy configurations, and as such they must be equated with a collective excitation of the field theory.  The simplest such objects are giant gravitons \cite{MST}. They are simple because they saturate a BPS bound, hence one can extrapolate between  weak and strong coupling limits.
 They come in two flavors: those that expand into the sphere, and those that expand into AdS \cite{Grisaru:2000zn,Hashimoto:2000zp}. These last ones are shown to be related to going to a Higgs configuration of 
 the ${\cal N}=4$ SYM theory on $S^3$ as is appropriate for radial quantization \cite{Hashimoto:2000zp} ( see also \cite{Btoy}). Giant gravitons are not only extended objects, but they come with  parameters determining how much R-charge they have. Their geometric position in the $AdS_5\times S^5$ geometry depends on these parameters. Hence they can be used a probes of the geometry of the extra dimensions. Identifying the position of these D-branes in the field theory dual leads to a tool that in principle lets us explore the geometry of higher dimensions. We should also notice that the solutions that correspond to D-branes with a fixed charge are not unique: they are moving along a periodic trajectory, so that once we find a configuration in the gravity theory, there is still a zero mode of the position of the D-brane along its trajectory. This is an angle as the trajectory is periodic. The angle direction together with the R-charge should be thought of as a pair of collective coordinates of the D-brane.
  
Now imagine having more than one such giant graviton at different positions. If these are D-branes, then we should be able to suspend strings between them. If the D-branes are not on top of each other, the strings should get a contribution to their mass that depends on the geometric distance between the branes. This distance depends both on the zero mode angle variables and the R-charge. If the answer is geometric in nature, the delocalized giant gravitons with fixed angular momentum can only average over all of the angle configurations and the effective Hamiltonian between these states will mix them with each other in a complicated way. If a formulation of the giant graviton states already has such collective coordinates in the description, the computation will be much simpler: all we need to do is compute the energy of the string and show that the result is geometric (local in the collective coordinates). 
 
The purpose of this paper is to show how this is done in a systematic way: how starting from field theory (particularly ${\cal N}=4 $ SYM and a $\BZ_2$ orbifold), one can build the giant graviton states with their collective coordinates, attach strings to them and compute the energies of the strings as functions of the collective coordinates. The approach that I take should be contrasted with the approach pursued by de Mello Koch et al. which has culminated in the results found in 
\cite{Koch:2011hb,deMelloKoch:2011ci} which took the other route. After writing the effective Hamiltonian and diagonalizing it, which took a herculean effort on combinatorics,  they were able to obtain an effective theory of harmonic oscillators, which they dubbed `spring field theory'. 

The repackaging in terms of collective coordinates does two things. First, it suggests an approximation to various correlators that simplifies computations substantially.  Without these approximations at the beginning, the amount of extra algebra that needs to be carried around is much worse. Secondly, the approximation suggests a new way to write the operators that thread the strings between the giants. This is algebraically cleaner than the original formulation of these states written down in \cite{BBFH}: in that basis each graviton has an exact $R$-charge and the strings stretched between them require a lot of combinatorial dexterity to manipulate ( see \cite{Bhattacharyya:2008xy} for example). The new prescription also makes the constraint associated to the gauge symmetries on the giant gravitons also easier to understand. 

 At the end of the computations, we find that the contribution to the masses of strings stretching between giants (at one loop order in the field theory) is a distance squared in the collective coordinate geometry derived from field theory.
This makes contact with its interpretation as a geometric D-brane, but in such setups one can also associate these masses to the Higgs mechanism associated to breaking an enhanced gauge symmetry of coincident D-branes down to a diagonal
group when we separate them. We see a Higgs mechanism for the emergent gauge symmetries realized geometrically in terms of collective coordinates of D-branes.  In the limit where the D-branes have small R-charge, one obtains a result that matches the dispersion relation of magnons in the ${\cal N}=4$ spin chain model \cite{SZ}.
 
The paper is organized as follows. In section \ref{sec:colcoor} the single giant graviton states and their collective coordinates are introduced. The giant gravitons are defined by the $\det(Z-\lambda)$ operators and $\lambda$ is their collective coordinate. The overlaps of these states and their effective action are computed.
 In section \ref{sec:emergent} I show how to write states where giant gravitons have strings attached. Here the results do not depend on the number of giant gravitons. The operators are written as products of determinants with traces, where in the trace one allows insertions of poles in $Z$, of the form $ (Z-\lambda_i)^{-1}$ for $\lambda_i$ one of the giant graviton coordinates. The Gauss law constraint for strings beginning and ending on different giant gravitons is implemented with these states.  In section \ref{sec:higgs} the mass of such strings suspended between giants is computed. In order to simplify the combinatorial problem, this is done in an orbifold field theory where I consider fractional brane giant gravitons and strings suspend between them. The contribution to the anomalous dimension of such a string is proportional to $g_{YM^2} |\lambda_1-\lambda_2|^2$. We show this way that the mass of $W$ bosons connecting between separated branes is proportional to a distance squared in these collective coordinates. Using this interpretation and relating it to the usual way that the Higgs mechanism is realized in brane constructions we get that the higgs condensate is proportional to $|\lambda_1-\lambda_2|$ which are collective coordinates of the state. In section \ref{sec:ost} I revisit the construction of the open spring theory setup of \cite{deMelloKoch:2011ci} in terms of this dynamics and I show that it results from a small modification of effective action introduced in section  section \ref{sec:colcoor} to include the one loop energies of the excited strings. The new equations of motion give rise to a system of harmonic oscillators in a first order formulation, and the motion of the open springs is identified with motion of hole defects on the eigenvalue droplet picture of \cite{Btoy}. In section \ref{sec:outlook} I make various remarks concerning applications of this work to various related setups that have been studied in the literature with other computational tools. The appendix contains important combinatorial identities and notation that are used in the main calculations of the paper.

\section{Giant graviton states and their collective coordinates}\label{sec:colcoor}

As described in the BBNS paper \cite{BBNS}, single giant graviton states are described by sub-determinants of a (holomorphic) field in ${\cal N}=4 SYM$ and some of its orbifolds. To simplify the discussion, we will consider for the time being only the ${\cal N}=4 $ SYM scenario for a $U(N)$ gauge group. The single giant graviton states will therefore be described by states of the form $\det_\ell Z$, where $\ell$ is the (exact quantized) angular momentum of the giant 
graviton. These are given by the following expression \footnote{This is a different normalization than in the original BBNS paper.}
\begin{equation}
\det\!_\ell Z = \frac 1 {N!}  { N\choose \ell} \epsilon_{i_1, \dots , i_\ell, i_{\ell+1} \dots, i_N}\epsilon^{j_1, \dots , j_\ell, i_{\ell+1} \dots, i_N} Z^{i_1}_{j_1} \dots Z^{i_\ell}_{j_\ell}
\end{equation}
These are gauge invariant operators made of a single $Z$, with $R$-charge $\ell$. The normalization for the R-charge are those for $SO(6)$, where it is the spin of the highest weight state of $SO(6)$, and where the vector representation has spin one. The energy of a state with angular momentum $\ell$ is $\ell$ (this reflects the fact that the dimension of a canonical scalar is $1$ in four dimensions).
In the notation of \cite{Bstudy} (see also the appendix), these states are given by
\begin{equation}
\det\!_\ell Z= \frac 1 {N!}  { N\choose \ell} \epsilon\epsilon( \underbrace{Z, \dots, Z}_{\ell \hbox{ times}}, 1, \dots 1)
\end{equation}
and using the results in that paper their normalization is given by
\begin{equation}
\langle \det\!_\ell \bar Z \det\!_\ell Z \rangle = \left[ \frac 1 {N!}  { N\choose \ell} \right]^2 N! (N-\ell) ! (\ell) ! (\ell)!= \frac{N !}{(N-\ell) !}
\end{equation}
This is the normalization according to the Zamolodchikov metric (with the insertion points of the correlators and the position factors stripped), which is also the normalization of the object $\det_\ell (Z)$ in a complex Gaussian matrix model. One can also think of this result as the normalization of a particular state of R-charge $\ell$ for the SYM compactified on $S^3\times \BR$ in the free field limit. This is the same normalization  as given by the Schur Polynomial basis of states \cite{CJR}, where such a state corresponds to a young diagram of one column with $\ell$ boxes. The simplified expression of the norm justifies the normalization coefficient. When $\ell=0$, it is just the usual normalization of the identity. For $\ell=N$ is the usual normalization of determinant correlators.

Notice moreover the following identity
\begin{equation}
\det(Z-\lambda) = \sum_{\ell=0}^N (-\lambda)^{N-\ell} \det\!_\ell(Z)
\end{equation}
which produces a generating series for the sub-determinants by expanding the determinant as a power series in $\lambda$ (the identity is a relatively easy combinatorial exercise left to the reader). At this stage, without motivation, the generating series can be considered as a formal object with a formal parameter.
If we consider $\lambda$ not as a formal parameter,  but as a complex number, we obtain a collection of states parametrized by $\lambda$.   We are supposed to interpret this object as a particular state in a Hilbert space of states of finite size. The Hilbert space of states is the subset of the Hilbert space of states of ${\cal N}=4 $ SYM on $S^3$ spanned by the sub-determinant operators (after using the operator-state correspondence), with the  norm as described above by the Zamolodchikov norm. The set of states parametrized by $\lambda$ spans the full Hilbert space of $N+1$ states from the truncation if we take arbitrary linear combinations of them, but the collection itself is only a one-parameter submanifold of the restricted Hilbert space with $N+1$ states and does not produce the most general quantum state spanned by the basis. We will consider these states parametrized by $\lambda$ as our giant graviton states, and $\lambda$ will be the collective coordinate for those states.

The natural reason to add these objects in this particular combination is due to the geometric interpretation of giant gravitons as holes in a two dimensional fermion droplet as described in \cite{Btoy}. To add a hole one considers a ground state wave function of $N+1$ fermions with the extra fermion located exactly at position $\lambda$. The Vandermonde determinant of the eigenvalues will have a term of the form $\prod(z_i -\lambda)$. If we say that there is a hole inserted at $\lambda$ then $\lambda$ is interpreted as just a parameter for a wave function for the remaining $N$ fermions and it does not have a probabilistic interpretation unto itself. The effect of the exclusion principle persists: no other eigenvalue can be located at $\lambda$. Seeing holes as D-branes is also natural from the study of the $c=1$ matrix model \cite{Douglas:2003up}. 

 Consider now the overlap between two such states, parametrized by $\lambda$, and $\tilde \lambda$, and given as follows:
\begin{equation}
\langle \det( \bar Z-\tilde \lambda^*)  \det( Z-\lambda) \rangle= \sum_{\ell=0}^N (\lambda \tilde \lambda^*)^{N-\ell}\frac{N !}{(N-\ell) !}=N!\sum_{\ell=0}^N (\lambda \tilde \lambda^*)^{\ell}\frac{1}{(\ell) !}
\end{equation} 
Upon conjugation, the adjoint matrix to the matrix of operators  $Z$, $Z^\dagger$ is called $\bar Z$ (this is the usual holomorphic convention in supersymmetric field theories), and the complex conjugate of $\lambda$ is $\lambda^*$ (we treat complex conjugation of numbers or parameters with a $*$ notation). 
The last entry on the equation above is obtained from rearranging the order of summation. What we see here is that by using the generating series for the determinant, we obtain an answer which is a truncated exponential.
By taking derivatives and evaluating at zero, we can recover completely the norm as described above for the set of states, as well as the orthogonality of operators of different charge. In this sense, the generating function encapsulates completely the combinatorial results for the norm of the different states, as well as the bound on the charge of a single giant graviton. 

For values of $\lambda\tilde \lambda^*$ that are sufficiently small (which we will address later), the truncated exponential and the exponential differ by a very small amount, so we can approximate the result as follows 
\begin{equation}
\langle \det( \bar Z-\tilde \lambda^*)  \det( Z-\lambda) \rangle \simeq N! \exp(\lambda \tilde \lambda^*) \label{eqn:overlapgiant}
\end{equation} 
For the time being, let us consider the approximate form as given above as if it were exact and let us try to understand what it means. What we want to do is interpret the algebraic expression above geometrically.

The idea is quite simple. Consider a coherent state for a single harmonic oscillator, described by a parameter $\alpha$, where the state is defined by $\ket{\alpha}=\exp(\alpha a^\dagger) \ket 0$, and $a^\dagger$ is the usual raising operator. The lowering operator is called $a$ and they satisfy the following  commutation relations $[a, a^\dagger]= 1$. If one computes
the overlap between two of these states (that have not been normalized to have unit norm), we get with a standard calculation that
\begin{equation}
\bra \beta \ket \alpha =  \bra 0 \exp( \beta^* a) \exp( \alpha a^\dagger) \ket 0 = \exp( \alpha \beta^*) \bra 0  \exp( \alpha a^\dagger) \exp( \beta^* a) \ket 0= \exp(\alpha \beta^*)\label{eqn:overlaphosc}
\end{equation} 
We now see that the equation \eqref{eqn:overlapgiant} is identical to equation \eqref{eqn:overlaphosc} if we identify $\lambda$ with $\alpha$ and $\tilde \lambda$ with $\beta$ except for a constant normalization factor of $N!$ which is independent of the parameters characterizing the giant graviton states. In this sense, the factor of $N!$ is trivial, since it can be absorbed in a constant normalization factor multiplying the definition of the state itself. What is important is that the result of the overlap calculation shows that the parameter $\lambda$ plays the same role for giant gravitons as the parameter $\alpha$ plays for coherent states in the harmonic oscillator. Moreover, the result shows that a lot of the
`hard' combinatorial algebra of the determinant contractions is encoded in the `simple' combinatorial algebra of raising and lowering operators for a harmonic oscillator. The most important thing is that $\lambda$ is a complex parameter, so it has both a norm and a phase, and both matter for the overlaps. The associated geometry for $\lambda$ is that of the complex plane: a manifold with two real dimensions. Now we will show that the complex plane will have the standard canonical (Kahler) form on it. We will also find the effective Hamiltonian for the giant gravitons, and we will show that it is that of an inverted Harmonic oscillator. The idea is to define normalized states
\begin{equation}
\ket \lambda \simeq  N_\lambda \det(Z-\lambda) \ket 0_{FT}
\end{equation}
where we have introduced the vacuum of the field theory on $S^3$ as $\ket 0_{FT}$. We also choose the states to be orthonormal, so $N_\lambda= [N! \exp(|\lambda|^2)]^{-1/2}$. In the rest of the paper we will abuse notation by ignoring the field theory vacuum $\ket 0_{FT}$ and we will call the states by the operators they are related to.

To define the canonical form for the parameter $\lambda$, we use the Berry phase associated to paths in the $\lambda$ complex plane by the connection
\begin{equation}
\int dt \bra {\lambda (t)} i \partial_t \ket{\lambda(t)} =\int dt \lim_{\tilde\lambda\to \lambda (t)} \bra {\tilde \lambda } i \partial_t \ket{\lambda(t)}
\end{equation}
We do the computation as follows
\begin{eqnarray}
 \bra {\tilde \lambda } i \partial_t \ket{\lambda(t)}&=& N! N_{\tilde\lambda}i  \partial_t\left( N_\lambda(t) \exp(\tilde \lambda \lambda(t) \right)\\
 &=&\left[ \exp( -|\tilde \lambda|^2 -| \lambda|^2) \right]^{1/2} \exp(\tilde \lambda \lambda(t) )\left( i \tilde \lambda \dot \lambda - i( 1/2) [ \lambda^* \dot \lambda+\dot \lambda^* \lambda] \right )
\end{eqnarray}
The terms in the last parenthesis come from the derivative of the overlap function and the normalization factor respectively. The reader should consult \cite{Shankar} for a more in-depth understanding of the Berry phase as related to coherent states.

When we take the limit we get that the Berry phase is given by
\begin{equation}
\frac i2 \int dt  [\lambda^* \dot \lambda - \dot \lambda^* \lambda] = \frac i 2\int \lambda^* d\lambda -  \lambda  d \lambda^*
\end{equation} 
The Berry phase integral should be interpreted as the form $\int p dq$ for a canonical pair of coordinates $p,q$ on a symplectic manifold. 
The curvature of the connection is $i d\lambda^* \wedge d\lambda$. This is the usual canonical form of the volume in the complex plane (depending on conventions, there is a factor of $2$ in the normalization of this volume form). The computation basically states that $\lambda$ and $\lambda^*$ are canonical conjugates of each other.

By the same token, we find that the Hamiltonian acting on states of charge $\ell$ is exactly $\ell$ (these are half-BPS states after all). This acts on the state $\det(Z-\lambda)$ as the first order differential operator $N-\lambda \partial_\lambda$, which when expressed in terms of the overlap \eqref{eqn:overlapgiant}  gives us $( N-\lambda \partial_\lambda) \exp( \lambda\tilde \lambda) = (N-\lambda \tilde\lambda)  \exp( \lambda\tilde \lambda) $. With a small amount of algebra we can put this together into an effective action form for the parameter $\lambda$ by using the Berry phase and the Schr\"odinger equation
\begin{equation}
S_{eff}= \int dt\left[\bra{\lambda} i \partial_t \ket{\lambda}- \bra{\lambda} H \ket{\lambda} \right] = \int dt \left[\frac i 2   (\lambda^* \dot \lambda - \dot \lambda^* \lambda) - ( N-\lambda \lambda^*)\right]\label{eq:effaction}
\end{equation}
The corresponding equations of motion are $i \dot \lambda+ \lambda=0$ and its complex conjugate. They are a first order version of the equations of motion of a harmonic oscillator. This way of thinking about the problem is the same intuition that was used in the derivation of the string action in the plane wave limit by Kruczenski \cite{Kruc} from considering the sigma model limit of spin chain Hamiltonians. Here we have derived an effective action for a giant graviton state. 

Notice that although at the beginning of the section the states were treated somewhat algebraically by combinatorial methods, by the end of the computation above we have trivialized the dynamics of the giant gravitons and their collective coordinate parameter $\lambda$ to be that of a  simple inverted harmonic oscillator.

There are two things that still need to be be addressed. The first one is: when does the approximation of the truncated exponential as a true exponential fail? The second issue to be addressed is how does the geometry of the complex plane relate to the AdS dual setup? This second one can not be answered without the first one.

To answer the first question, let us consider the norm squared of a state 
\begin{equation}
\langle \det( \bar Z- \lambda^*)  \det( Z-\lambda) \rangle = N!\sum_{\ell=0}^N (|\lambda|^2)^{\ell}\frac{1}{(\ell) !}
\end{equation}
and compare it to $N! \exp(|\lambda|^2)$. This is, let us evaluate the partial sum
\begin{equation}
1-\exp(-|\lambda|^2)\sum_{\ell=0}^N (|\lambda|^2)^{\ell}\frac{1}{(\ell) !}= \exp(-|\lambda|^2) \sum_{\ell=N+1}^\infty(|\lambda|^2)^{\ell}\frac{1}{(\ell) !}
\end{equation}
The individual entries are given by $(|\lambda|^2)^{\ell}/ {(\ell) !}\exp(-|\lambda|^2)$ and these can be considered as probabilities for a Poisson distribution with parameter $|\lambda|^2$. At large $|\lambda|$ (which is the limit we are interested in) the Poisson distribution becomes approximately Gaussian with the maximum probability peak located at $\ell \simeq|\lambda|^2$, so  for these terms  $\ell!$ can be approximated by  Stirling's approximation
\begin{equation}
(|\lambda|^2)^{\ell}/ {(\ell) !}\simeq \exp( \ell \log|\lambda|^2 - \ell \log\ell+\ell)
\end{equation}
In the saddle point limit, where $\ell$ is treated as a continuous parameter we find that the maximum is located at $\ell_0 = |\lambda|^2$, and that the Gaussian approximation is given by
\begin{equation}
(|\lambda|^2)^{\ell}/ {(\ell) !} \simeq \exp |\lambda|^2 \exp(- (\ell-\ell_0)^2/ \ell_0)
\end{equation}
Now, the partial sum of the terms that are omitted is approximated by the integral
\begin{equation}
\int_{\ell=N+1}^\infty \exp(- (\ell-\ell_0)^2/ \ell_0)
\end{equation}
which is a cumulative function for a tail of the Gaussian probability distribution centered at $\ell_0=|\lambda|^2$ with standard deviation $\sigma=\ell_0/ 2$. To $k \sigma$ accuracy, we need that 
that $N>\ell_0$, so that we are above the bump,  and that $(N-\ell_0)^2 > k\sigma \simeq k \ell_0$, so that  $\ell_0<N-O(\sqrt{N})$. This is, $|\lambda|^2$ can be as large as $N$ with a small correction of order $\sqrt{N}$ depending on the precision we want. The residue is exponentially suppressed in $N$ and thus is a non-perturbative correction to the norm. In most circumstances we can throw away these terms with impunity as they will not affect $1/N$ corrections.

If we consider the Hamiltonian we found, $H_{eft} = N-|\lambda|^2$, the condition for the coherent state to make sense as being approximated by a classical state in an inverted harmonic oscillator is that it has positive energy $H_{eff}>O (\sqrt N)$. This is natural considering that all excitations have positive energies in the ${\cal N}=4 $ SYM theory, due to unitarity of the associated conformal field theory. Thus, one can say that the allowed phase space is constrained by unitarity \footnote{This is a field theory analog  to the observation of \cite{Caldarelli:2004mz,Milanesi:2005tp} that unitarity is related to the absence of closed time-like curves in the corresponding supergravity droplet solutions, and that fermion density and this relates to the observation that in the free fermion droplet picture  the hole density functions  are bounded by the density that saturates the Pauli exclusion  principle}. 

Thus, we find that the allowed set of states defined by the parameter $\lambda$ is a disk centered at zero with radius $R= \sqrt N$. The disk in the complex plane at the center is the fermion droplet picture of the half-BPS states that was developed in \cite{Btoy}. The giant graviton coherent states are circular holes in this Fermi-liquid distribution with a fixed area of order one (more precisely $\pi$) while the area of the droplet is 
$ \pi N$.  Also, we find that the approximations for calculating the norms break down for configurations of energy of order $\sqrt N$ coming from above. For the case of trace operators, the planar approximation to calculations for norms breaks down for energies of order $\sqrt N$ from below \cite{KPSS,Btoy}. 
 In the supergravity theory, the same picture of droplets is available \cite{LLM} and a complete construction of the corresponding supergravity solutions for the most general such droplet configuration has been found. Studying such general configurations is beyond the scope of the present paper.

Notice also that we can rescale the droplet so that it has radius one. This is done by choosing a variable $\xi = \lambda/\sqrt N$. In terms of the variable $\xi$ we have that the allowed disk in the complex plane is of radius one, while the action is given by 
\begin{equation}
S_{eff}= N \int dt \left[\frac i 2   (\xi^* \dot \xi - \dot \xi^* \xi) - (1-\xi\xi^*)\right]
\end{equation}
What is important is the factor of $N$ appearing as a prefactor in the action. This factor of $N$ indicates that in a planar diagram interpretation of matrix mechanics, that the object we are considering is a D-brane. After all, the effective planar diagram string coupling constant is $g_{eff}^2\simeq 1/N^2$ \cite{'tHooft}, and the tension of a D-brane is of order $1/g\simeq N$. The parameter $\xi$ is the position of the D-brane in the disk geometry. 

For multiple giant gravitons the answer should be essentially the same. The states should be of the form $\det\left[( Z-\lambda_1)(Z-\lambda_2)\dots (Z-\lambda_k)\right]$, and for each such $\lambda_i$ a similar
action as above would be generated.  
Working out the full details of the combinatorics that allows this to happen will be postponed for another occasion. It should follow directly from the Schur basis  results of \cite{CJR} after the appropriate repackaging in 
the generating series is done. It should be noted that in computing overlaps as we did above, that the different $\lambda_i$ have a permutation symmetry. This symmetry should be implemented by stating that the parameters $\lambda_i$ are similar to the way we think of the eigenvalues of $Z$ as fermionic coordinates, since after all, the giant gravitons are thought of as  holes in the fermion droplet picture. The holes have the same underlying permutation symmetry as the objects that they represent a vacancy from, which in this case is the constituent fermions of the fermion droplets. When the $\lambda_i$ are sufficiently different from each other (and not too many) there are no statistics to worry about as the states are clearly distinguishable and the Pauli exclusion principle says nothing: no state has any sizable probability of being doubly occupied. This is a consequence of the fact that coherent state overlaps are exponentially suppressed for large distances.

\section{Attaching strings and emergent gauge symmetries}\label{sec:emergent}

The recipe for attaching strings to a single subdeterminant operator was proposed in \cite{BHLN}, following closely the set of states identified in \cite{BHK} for the conifold field theory. In the notation of \cite{Bstudy} such states are identified (for a single string excitation) as
\begin{equation}
\epsilon\epsilon( \underbrace{Z, \dots Z}_{\ell -1 \hbox{ terms}}, W_1, 1, \dots, 1)
\end{equation}
and for many excitations as 
\begin{equation}
\epsilon\epsilon( \underbrace{Z, \dots Z}_{\ell -s \hbox{ terms}}, W_1,W_2, \dots W_s, 1, \dots, 1)\label{eq:words}
\end{equation}
where the $W_s$ are words in the ${\cal N}=4 $ SYM fields. The approximate orthogonality of the states with different numbers of excitations $s<< N$ was shown in \cite{Bstudy}.

Let us specialize to maximal giants, where we have $\ell =N$. We will now rewrite the list of states in a slightly different form that makes them slightly more tractable. The idea is to consider the operator
\begin{equation}
O(\kappa_1, \dots \kappa_s) =\det\left( Z+ \sum_{i=1}^s \kappa_i W_i\right) = \frac 1 {N!} \epsilon\epsilon\left( Z+ \sum_{i=1}^s \kappa_i W_i, \dots, Z+ \sum_{i=1}^s \kappa_i W_i \right)
\end{equation} 
where the $\kappa_i$ are formal parameters. It is easy to see that terms similar to those in expression \eqref{eq:words} are generated by 
the formal expansion. Indeed, we can check easily that 
\begin{equation}
\frac 1 {N!}\epsilon\epsilon( \underbrace{Z, \dots Z}_{N-s \hbox{ terms}}, W_1,W_2, \dots W_s) \propto \partial_{\kappa_1} \dots \partial_{\kappa_s} O(\kappa_1, \dots \kappa_s) |_{\kappa_1=\dots =\kappa_s=0} 
\end{equation}
This follows because the $\epsilon\epsilon$ symbol is multilinear in all of its entries, and it is also a symmetric function of the entries. In the case that the words $W_i$ are fermionic, the corresponding formal parameters $\kappa_i$ are to be treated as Grassman parameters.  
The proportionality constant is $[N(N-1) \dots (N-s+1)]^{-1}$. The normalization comes from the combinatorics of choosing on which entries in the symbol the derivatives act. No two derivatives can act on the same symbol entry because the matrix function  $Z+ \sum_{i=1}^s \kappa_i W_i$ has no double derivatives with respect to the $\kappa_i$. 
The next simple observation is that we can factor the $\det(Z)$ from $O(\kappa_1, \dots \kappa_s)$. This is, we can formally write
\begin{equation}
O(\kappa_1, \dots \kappa_s)= \det(Z) \det\left( 1+ \sum_{i=1}^s Z^{-1} \kappa_i W_i\right)
\end{equation}
where we assume that $Z^{-1}$ is a well defined matrix. Although the reader might be uncomfortable with the introduction of $Z^{-1}$ when it is operator valued, algebraically the full expression is a polynomial in the $Z$ entries. The manipulation at the formal level is a convenient way to understand how the string ends will be generated.

This immediately leads to the following identity
\begin{equation}
\frac 1 {N!}\epsilon\epsilon( \underbrace{Z, \dots Z}_{N-s \hbox{ terms}}, W_1,W_2, \dots W_s)= \det(Z) 
\frac 1 {N!}\epsilon\epsilon( \underbrace{1, \dots 1}_{N-s \hbox{ terms}}, Z^{-1} W_1,Z^{-1} W_2, \dots Z^{-1}  W_s)
\end{equation}
Expressions like $\epsilon\epsilon( \underbrace{1, \dots 1}_{N-s \hbox{ terms}}, \tilde  W_1,\tilde W_2, \dots, \tilde W_s)$ have a nice expansion in terms of multi-traces of all the orderings of the entries of $\tilde W_i$ \cite{Bstudy}. 
The simplest example is that for $s=1$. In that case, we find that 
\begin{equation}
 O(\kappa_1) = \det(Z) \det( 1+ \kappa_1 Z^{-1} W_1) = \det(Z)\left( 1 + \Tr(\kappa_1 Z^{-1} W_1) + O(\kappa_1^2)\right)
\end{equation}
So that the operator of interest is given by
\begin{equation}
\partial_{\kappa_1}  O(\kappa_1)|_{\kappa_1= 0} = \det(Z) \Tr(  Z^{-1} W_1)
\end{equation}
If we want a formula  for the expansion in  a general case, we use the following expression
\begin{eqnarray}
 O(\kappa_1, \dots \kappa_s)&=& \det(Z) \exp\left[\Tr\left( \log\left [ 1+ \sum_{i=1}^s Z^{-1} \kappa_i W_i\right]\right)\right]
 \\ &=& \det(Z) \exp\left[\Tr\left( \sum_{t=1}^\infty \frac{(-1)^{t+1}}{ t} \left [  \sum_{i=1}^s Z^{-1} \kappa_i W_i\right]^t\right)\right]
 \end{eqnarray}
and to compute the derivatives at zero that we want,  we can truncate the sum in the exponential to order $t=s$. The same is true with the Taylor series of the exponential itself. This series reproduces all we need. Notice that for terms with $t$ matrices inside a trace, there is a $1/t$ in front of it. This factor counts the identification of traces under cyclic permutations, so that each possible ordering of the matrices appears only once in the trace, with coefficient equal to one, and we do not add the cyclic permutations of the object.

When we expand we realize that we are allowed to use inverse powers of $Z$ inside traces. But each of these always show up sandwiched between the $W_i$. This is, expressions like $\Tr( Z^{-1} W_1) \Tr(Z^{-1} W_2)$ are allowed, but not
$\Tr(Z^{-2} W_1) \Tr( W_2)$. Similarly $\Tr( Z^{-1} W_1 Z^{-1} W_2)$ is allowed, but $\Tr( Z^{-2} W_1 W_2)$ is not.
An equally good expansion could have been gotten if we used
\begin{equation}
\det\left( Z+ \sum_{i=1}^s \kappa_i W_i\right) = \det\left( Z^\gamma_1 \left(1 + \sum_{i=1}^s Z^{-\gamma_1} \kappa_i W_i Z^{-\gamma_2} \right)Z^{\gamma_2} \right ) \label{eq:redundancy}
\end{equation}
for $\gamma_1+\gamma_2=1$ fixed constants.  Then we would learn that the expansion would involve traces of words of the form $Z^{-\gamma_1} W_i Z^{-\gamma_2}$ and then when we combine them they appear always together as $Z^{-\gamma_2} Z^{-\gamma_1}= Z^{-1}$ just as above. The idea of introducing a redundancy by hand with a constraint allows one to think of the phenomenon as an emergent gauge symmetry, in this case on the worldvolume of the giant gravitons,  and it will play the same role as the combinatorics described in \cite{BBFH}. 

Now that we understand how to generate the configurations with $s$ strings attached to a maximal giant graviton, let us consider how we do this for a non-maximal giant. The answer should be pretty obvious: everywhere where $Z$ appears, we use $(Z-\lambda)$ instead. This includes in the denominators inside the traces. Thus, our giant gravitons with strings attached will be of the form
\begin{eqnarray}
\CO (W_1, \lambda)&=& \det(Z-\lambda) \Tr\left(\frac 1{Z-\lambda} W_1\right)\label{eq:examplewords}\\
\CO (W_1, W_2, \lambda)&=& \det(Z-\lambda)\left[ \Tr\left(\frac 1{Z-\lambda} W_1\right) \Tr\left(\frac 1{Z-\lambda} W_2\right)- \Tr\left( \frac 1{Z-\lambda} W_1\frac 1{Z-\lambda} W_2\right)\right] \nonumber
\end{eqnarray}
and these generalize easily to more words.

The idea is now to think about how to do this same trick for multiple giants. As described in the previous section, the basic object we want is given by
\begin{equation}
\CO(\lambda_1, \dots \lambda_k)= \det\left ( (Z-\lambda_1) \dots (Z-\lambda_k)\right)
\end{equation}
The idea is that now for each $\lambda_i$ we introduce the same factors of $\gamma_1^i+\gamma_2^i =1$ as in equation \eqref{eq:redundancy}, but the label $i$ attached to the $\gamma$ indicates that each is to be treated independently of the others.

Now, a word stretching between the giants $i, j$ will have a prefactor  $(Z-\lambda_i)^{-\gamma^i_1}$ and a post factor of 
$(Z-\lambda_j)^{-\gamma_2^j}$. A typical example of a full operator would be
\begin{eqnarray}
&\det[(Z-\lambda_1)(Z-\lambda_2)] \Tr \left[(Z-\lambda_1)^{-\gamma_1^1} W_1 (Z-\lambda_2)^{-\gamma_2^2}  (Z-\lambda_2)^{-\gamma_1^2}W_2 (Z-\lambda_1)^{-\gamma_2^1} \right]&\\
&= \quad \det[(Z-\lambda_1)(Z-\lambda_2)]  \Tr[ (Z-\lambda_1)^{-1} W_1 (Z-\lambda_2)^{-1} W_2 ] &
\end{eqnarray}

To understand how these operators are built, it is convenient to think of $Z$ as a diagonal matrix with eigenvalues $z_r$, and to consider the $r,s$ entries of the worded $W_1$ etc. The entries will have the form
\begin{equation}
 {(z_r-\lambda_i)^{-\gamma_1^i}}  (W_1)_{rs} (z_s-\lambda_j)^{-\gamma_2^j} 
\end{equation}
These can generate a branch cut in the $z_r$ and $z_s$ variables. However, we have that the operator should be a polynomial in the 
$Z$ fields, and hence polynomial in the $z_r$. The allowed operators are therefore free of branch cuts. Each time a word starts on brane $i$ and ends on brane $j$ we seem to introduce a branch cut at $z_r= \lambda_j$.
To cancel the branch cut we need another string to start at brane $j$ for it to be paired with. Then we can use the $\gamma_1+\gamma_2=1$ constraint and the possible 
branch cut is eliminated. The result is that we get a pole at $z_r=\lambda_j$. Such a pole is allowed, because the determinant has a zero at that location. However, if we get a double pole at $z= \lambda_j$, we are in trouble (such objects are non-normalizable wave functions in the fermion picture). Also, gauge invariance of the original system requires us to sum over eigenvalues at each such insertion. Thus we see that we naturally get matrix multiplications that need to be closed into traces. The (in principle strange) combinatorics that arise in the traces, as in equation  \eqref{eq:examplewords} make sure that all the possible double poles at $\lambda_i$ ( and higher poles in general) are cancelled between the different trace combinations. This idea of using inverses of matrices has also been investigated in a more restricted setting in \cite{CCS}, where the calculations were done for concentric droplet configurations in a saddle point approximation.

A simple consequence of this investigation is that we get a constraint: for each string beginning on brane $i$, another string must end on it. This is the Gauss' law constraint, and it shows that we have an emergent  $U(1)$ gauge symmetry for each $\lambda_i$. The equation that 
determines this constraint is $\partial_{\gamma_1} X =0$, when using $\gamma_2=1-\gamma_1$, for the different $\gamma^i$ pairs. 
 In the end, this is easier to write down than the prescription in terms of Young tableaux developed in \cite{BBFH}, but it has the same information content.

There is one more thing we can check: we need to check that this procedure saturates all possibilities. Let us suppose that a perverse demon gives us a word $W_1$ and that it requires it to have 
a prefactor not just for brane $i$, but also for brane $i '$. This might suggest that there are also three-pronged objects in the theory that are not just strings stretching between two branes, which only have two ends. Thus, such a word would be accompanied by  a prefactor on the left of the form
\begin{equation}
(Z-\lambda_i) ^{-\gamma_1^i} (Z-\lambda_i' ) ^{-\gamma_1^{i'}} W_1
\end{equation}
and the same reasoning as above would require us to pair it with a word with a post-factor given by
\begin{equation}
W' (Z-\lambda_i) ^{-\gamma_2^i} (Z-\lambda_i' ) ^{-\gamma_2^{i'}}
\end{equation}
so that the end result is single valued on the eigenvalues. 
The single valued combination would look as 
\begin{eqnarray}
&&W' (Z-\lambda_i) ^{-\gamma_2^i} (Z-\lambda_{i'} ) ^{-\gamma_2^{i'}}(Z-\lambda_i) ^{-\gamma_1^i} (Z-\lambda_{i'} ) ^{-\gamma_1^{i'}} W_1\\
&=& W' (Z-\lambda_i)^{-1}(Z-\lambda_{i'})^{-1} W_1\\
&=& W' \frac{1}{\lambda_i -\lambda_{i'}} \left( \frac 1 {Z-\lambda_i }- \frac 1{Z-\lambda_{i'}}\right) W_1
\end{eqnarray}
where in the last part we have used a summation by partial fractions identity.  Remember that in this discussion $\lambda_i$ and $\lambda_{i'}$ are parameters (numbers), so we can manipulate them as such. This shows that the end result is a linear combination of strings stretching between brane $i$ and other branes, and strings stretching between brane $i'$ and other branes, and that there is no doubly pronged initial word starting at both brane $i$ and $i'$ simultaneously. This is, the strings only stretch between pairs of branes as one usually imagines them to do. Similar partial fraction identities   get rid not just of double-pronged prefactor objects, but of multiple-pronged prefactor objects in general. The absence of double poles at each $\lambda_i$ gets rid of repeated ends on the same brane. The end result is that one can count the string states stretching between branes in the same way one counts with matrix entries. The reader should compare this with \cite{deMelloKoch:2012ck} where such a counting is shown to be true by purely combinatorial methods.

At this stage we have accomplished two main goals of this paper: first, to obtain a legitimate collective coordinate for giant gravitons (and deriving its effective action) and secondly to find a simple description of states with strings attached to giant gravitons.
The next task is to do a computation of the mass of such strings in a simple example to show the simplifying power of these techniques.

\section{Masses of strings stretched between distant giants and the Higgs mechanism for emergent gauge symmetries} \label{sec:higgs}

The first thing the reader should notice is that we have so far been able to do a very detailed analysis of overlaps for a single giant graviton state, but we have only sketched what needs to be done for multiple giant gravitons. What I want to do in this section is repackage the single giant graviton results so that they can be used in a multiple giant graviton setting essentially without modification. 

The idea is to consider a ${\mathbb Z}_2$ orbifold field theory of ${\cal N}=4 $ SYM, so that we have an $N=2$ SYM theory for D-3 branes at the tip of the the $A_1$ singularity $\BC^2/\BZ_2$. The idea of using orbifolds to try to understand giant graviton states is part of the original BBNS argumentation \cite{BBNS}, so we will follow this same route.

The simplest way to deal with such theories is in terms of a quiver diagram obtained by the procedure of Douglas and Moore \cite{Douglas:1996sw}. A convenient way to repackage this information is in terms of of a cross-product algebra \cite{BL}, and in general one can associate a quiver algebra with relations to any supersymmetric field theory \cite{Brev} (see also \cite{BRomo} for how to do this in theories with Chern-Simons couplings, and where some of these ideas are elaborated in detail). Instead of writing the $\BZ_2$ quiver itself, we will consider  the two projectors associated to the nodes of the quiver $\pi_1, \pi_2$, and we will diagonalize them so that 
\begin{equation}
\pi_1 = \begin{pmatrix} {\bf 1} &0\\0&0\end{pmatrix}, \quad \pi_2= \begin{pmatrix}  0&0\\0&{\bf 1}\end{pmatrix}
\end{equation}
The usual arrows of the quiver will be block entries between the nodes. We choose the $\BZ_2$ action to act as $X\to -X, Y\to -Y, Z\to Z$, so that the block matrices for the quiver are given by
\begin{equation}
X= \begin{pmatrix} 0 & X_{12}\\ X_{21} &0\end{pmatrix}, \quad Y=\begin{pmatrix} 0 & Y_{12}\\ Y_{21} &0\end{pmatrix}, \quad Z= \begin{pmatrix} Z_1 &0\\0&Z_2\end{pmatrix}
\end{equation}
and it is clear that we can recover the different fields by sandwiching $X,Y,Z$ between projectors $\pi_i$. For example $Z_1\simeq \pi_1 Z \pi_1$ etc. Similarly, a trace performed on the node $1$ of the quiver would have the general form $\Tr( \pi_1 f(X,Y,Z ))$ where $f$ is a general word in the quiver, including possible additional insertions of the $\pi_i$. Thus for example $\Tr(Z_1^k)=\Tr(Z^k \pi_1)$, etc.

The important thing is that the orbifold projection does not change the form of the superpotential. It is still given by the $\Tr(XYZ-XZY)$ form of the ${\cal N}=4 $ SYM, which can be expanded into components by using the matrix representation for $X,Y,Z$ in matrix form described above. Also, we can apply the same ideas to derivatives with respect to the letters of the words. Thus, we also have that 
\begin{equation}
\partial_X= \begin{pmatrix} 0 &\partial _{X_{21}}\\ \partial_{X_{12}} &0\end{pmatrix}, \quad \partial_Y=\begin{pmatrix} 0 & \partial_{Y_{21}}\\ \partial_{Y_{12}} &0\end{pmatrix}, \quad \partial_Z= \begin{pmatrix} \partial_{Z_1} &0\\0&\partial_{Z_2}\end{pmatrix}
\end{equation}
Notice that derivatives with respect to an arrow that goes from node $i$ to node $j$ is an operation that removes such an arrow. Thus, it has the opposite quantum numbers and as such the derivatives with respect to such objects run in the opposite direction in the quiver, from node $j$ to $i$. 

Although at first it might seem that this is academic, since we are complicating the details of the field theory, our objective is to find a simple situation where one can talk about two such giant gravitons and use the combinatorics for a single giant graviton effectively.
The basic two giant graviton operator we want is the state
\begin{equation}
\CO(\lambda_1, \lambda_2) =\det( Z_1- \lambda_1) \det(Z_2-\lambda_2)
\end{equation}
 Because it is factorized in the $Z_1$ and $Z_2$ degrees of freedom, the results from \eqref{eqn:overlapgiant} just get doubled without any extra work. This is, if we use unnormalized operators then
 \begin{equation}
 \bra{\tilde\lambda_1,\tilde  \lambda_2}\ket{\lambda_1,\lambda_2}= (N!)^2 \exp( \tilde \lambda^*_1\lambda_1+\tilde\lambda_2^*\lambda_2)
 \end{equation}
 Also, since the maximal value of $\lambda$ is $\sqrt N$, we find that the parameters $\lambda_i$ should be treated as being of order $\sqrt N$ as far as large $N$ counting goes.
 
 The two giant gravitons are distinguishable, and one should think of them as fractional branes themselves, whereas $\det(Z-\lambda)$ should be a regular giant graviton. A way to think about this is that the operator $\det(Z_1-\lambda_1)$ sources twisted sector states (these are of the form $\Tr(f(X,Y,Z) (\pi_1-\pi_2)$, whereas untwisted sectors are just traces $\Tr(f(X,Y,Z)(\pi_1+\pi_2))$ see \cite{BLeigh,BL} for more details on how to describe  these states). This is easiest to interpret if we expand $\det(Z_1-\lambda)$ in terms of the traces of powers of $Z_1$. Such traces are linear combinations with a non-zero overlap to twisted sector closed string states. 
 
 Now, we want to add a string stretching from brane $1$ to brane $2$, and also one such string stretching in the opposite direction. We can choose these to be given by the operator
 \begin{equation}
\CO(\lambda_1, \lambda_2, Y_{12}, Z_{21}) = \det( Z_1- \lambda_1) \det(Z_2-\lambda_2) \Tr \left[\frac 1{ Z_1-\lambda_1} Y_{12} \frac 1{ Z_2-\lambda_2} X_{21}\right]
 \end{equation}
The choice of taking a different letter in the $12$ direction versus in the opposite direction is that 
it makes it much simpler to separate the contributions to the energy from both strings. The result will not matter in regards to how we made this choice. 

The idea now is to copy the result for the one lop dilatation operator of ${\cal N}=4 $ SYM in abstract form for the $Y,Z$ degrees of freedom as given by \cite{BKS}
\begin{equation}
H\propto g_{YM}^2 \Tr( [Y,Z][\partial_Z, \partial_Y])
\end{equation}
and to apply to the system above. There is a similar contribution for the $X,Z$ Hamiltonian. Again, we will look at this piece alone in order to isolate the contribution of the string $Y$. The main problem we face in evaluating these energies in general is that we can only consider states that are allowed by the Gauss law constraint. Thus we never get just one string contributing to the energy, but the sum over the different strings.  Here, we can isolate the contribution and use symmetry to show that the same result will hold for the $X$ letter. Since we only have one $Y_{12}$ letter, the effective Hamiltonian simplifies to
\begin{eqnarray}
H&\propto& g_{YM}^2 \Tr\left( Y_{12} Z_2 \partial_{Z_2} \partial_{Y_{12}}-
Z_1Y_{12} \partial_{Z_2} \partial_{Y_{12}}\right.\\
&&-  \left. Y_{12} Z_2\partial_{Y_{12}}\partial_{Z_1}
+Z_1 Y_{12} \partial_{Y_{12}}\partial_{Z_1}\right)
\end{eqnarray}
Now, we need a couple of results to be able to take the derivatives. For the determinant, we have that the matrix derivative is
\begin{equation}
(\partial_Z)^a_b \det(Z-\lambda) = \det(Z-\lambda) \left(\frac 1{Z-\lambda}\right)^a_b
\end{equation}
And for expressions of the form
\begin{equation}
(\partial_Z)^a_b\left( \frac 1{Z-\lambda}\right)^c_d =
- \left(\frac 1{Z-\lambda}\right) ^c _b  \left(\frac 1{Z-\lambda}\right) ^a _d  
\end{equation}
Thus these derivatives open up a trace into two traces.
 
The first term acting on the state gives us two terms. We will concentrate on the one where the action of the derivative with respect to $Z$ is in the determinant. We find that the result is
\begin{equation}
g_{YM}^2 \Tr\left( Y_{12} Z_2 \partial_{Z_2} \partial_{Y_{12}}\right) \CO(\lambda_1, \lambda_2, Y_{12}, Z_{21}) = g_{YM}^2 \CO(\lambda_1,\lambda_2) \Tr \left[\frac 1{ Z_1-\lambda_1} Y_{12} \frac {Z_2}{ (Z_2-\lambda_2)^2} X_{21}\right]
\end{equation}
whereas the second term of $H$ acting on the same state gives us
\begin{equation}
-g_{YM}^2 \Tr\left( Z_1Y_{12}  \partial_{Z_2} \partial_{Y_{12}}\right) \CO(\lambda_1, \lambda_2, Y_{12}, Z_{21}) = -g_{YM}^2 \CO(\lambda_1,\lambda_2) \Tr \left[\frac {Z_1}{ Z_1-\lambda_1} Y_{12} \frac {1}{ (Z_2-\lambda_2)^2} X_{21}\right]
\end{equation}
We see that we have generated double pole terms. As discussed previously, these have to be absent in general: they are cancelled in the muti-trace expansion. It must be the case that the
other extra terms where $\partial_Z$ acts on $1/(Z-\lambda)$ have to cancel these. Now let us 
replace $Z_1\to Z_1 -\lambda_1 +\lambda_1$ and similar for $Z_2$ in the numerators. We see that we generate traces of the form
\begin{equation}
\Tr\left[ Y_{12} \frac {1}{ (Z_2-\lambda_2)^2} X_{21}\right]
\end{equation}
that have no single poles at $Z_1=\lambda_1$. 
These  are interpreted as single strings that stretch from the second brane to itself. These should be suppressed by $1/N$ when we do an overlap, according to the general observations in \cite{Bstudy}. We will assume that these can be dropped. Thus, for the pieces that survive,  we find a result proportional to 
\begin{equation}
\lambda_1 \Tr\left[\frac 1 { Z_1-\lambda_1}Y_{12} \frac {1}{ (Z_2-\lambda_2)^2} X_{21}\right]
\end{equation}
The full expression of the double pole term for a single trace in $\lambda_2$ is therefore
\begin{equation}
g_{YM}^2 (\lambda_2-\lambda_1) \Tr\left[\frac 1 { Z_1-\lambda_1}Y_{12} \frac {1}{ (Z_2-\lambda_2)^2} X_{21}\right]
\end{equation}
and the double trace then generated must by force have the correct combinatorics to cancel the double poles in the eigenvalues of $Z$ as those we have seen in expressions like equation \eqref{eq:examplewords}. To put it into the form of equation \eqref{eq:examplewords}, we take the 
two words to be $\tilde W_1= 1$ and the composite word $\tilde W_2= X_{21}(Z_1-\lambda_1)^{-1}Y_{12} $ for example. 

The other terms where we take derivative with respect to $\partial_{Z_1}$ similarly generate double poles in $\lambda_1$ which are proportional to $\lambda_1-\lambda_2$ and give a similar expression as above.

The single pole terms that are generated in the process are similar to what we started with at the beginning and in principle give a contribution to the anomalous dimension which is of order $g_{YM}^2$ on the diagonal, rather  than proportional to the t'Hooft coupling $g_{YM}^2 N$, so we should think of them as subleading in a planar expansion and we can also drop them.

The last thing we need to do is the overlap of this state with the original state, so that when we evaluate the Hamiltonian it is of the form
\begin{equation}
\bra{\lambda_1, \lambda_2, Y_{12}, X_{21}} H \ket{\lambda_1, \lambda_2, Y_{12}, X_{21}}
\end{equation}
where the states are normalized to have unit norm. 
To do this, we use the results from generating functions in the appendix, in particular equations \eqref{eq:main1}, \eqref{eq:main2}  to find that at leading order in $N$ the anomalous dimension is proportional to
\begin{equation}
g_{YM}^2 (\bar \lambda_1-\bar\lambda_2) (\lambda_1-\lambda_2)= g_{YM}^2 |\lambda_1-\lambda_2|^2 \label{eq:stringstretch}
\end{equation}
The end result is remarkably simple and it just involves the distance squared in the complex plane between the two parameters $\lambda_1$ and $\lambda_2$. 

Notice that the result is quadratic in the differences of the $\lambda_i$, and that it is invariant under simultaneous rotations of the $\lambda_i$ by a common phase. 
This requirement just states that the end result preserved the $R$-charge of rotations in $Z$, which is inherited as an unbroken geometric symmetry from the parent ${\cal N}=4 $ SYM theory. It is a consistency check of the formalism. Notice also that the result is local in the $\lambda$ variables. This is because $\bra{\tilde \lambda_1, \tilde \lambda_2, Y_{12}, X_{21}} H \ket{\lambda_1, \lambda_2, Y_{12}, X_{21}}$ are exponentially suppressed when the $\lambda$ are very different from each other, thus the effective Hamiltonian does not induce non-local mixing between states with roughly the same energy.

The expression also has the correct scaling to be compatible with the t'Hooft expansion. This justifies our expectation to be able to drop various terms in the expansion before because they would be subleading in $N$. In terms of our normalized $\xi$ variables, this is given by
\begin{equation}
 g_{YM}^2 N |\xi_1-\xi_2|^2
\end{equation}
which makes the 't Hooft scaling manifest.
It is also clear that the same reasoning will go for the letter $X_{21}$, but using the effective Hamiltonian 
that has $\Tr[ X,Z][\partial_Z,\partial_X]$ instead.

The way we are supposed to interpret this fact is straightforward. When we separate the giant gravitons in the droplet plane, the off-diagonal strings stretching between them get a mass proportional to $g_{YM}^2 |\lambda_1-\lambda_2|^2$, this is apart from their $R$ charge contribution to the energy (equal to $1$). If we interpret this energy as resulting from momentum quantization plus a mass, the full energy squared is of order $p^2+m^2\simeq 1+O(g_{YM}^2 |\lambda_1-\lambda_2|^2)$. The correct result should be the  square root of such an expression,  which to leading order in a Taylor seres $g_{YM}^2$  gives the same result as above. However, for large 't Hooft coupling we would expect the 
square root to be dominated by the distance itself. This would require resumming anomalous dimensions to all orders in the planar approximation.  The reader should notice the following: we can take the limit where the parameters $|\lambda_i|\to \sqrt{N}$ where they approach the edge of the droplet. In that limit the giant gravitons become giant gravitons in the plane-wave limit \cite{LLM} and we should be able to compare the system with the corresponding limit in field theory \cite{BMN}. 
 In that geometric limit 
the expression for the energy of the  string stretching between giant gravitons is identical to the description of a BMN operator closed string states in the planar limit \cite{BCV}, not just as a number, but also in terms of the geometric interpretation.  The \cite{BCV} calculation is done as a saddle point in a model of commuting matrices  with off-diagonal excitations (the model of commuting matrices was suggested in \cite{Berenstein:2005aa} as a theory of the geometry of the $S^5$ geometry. In a sense, the calculation in \cite{BCV} it is an uncontrolled approximation because we don't know to what extent the assumptions about commuting matrices are valid at strong coupling. That calculation matched the anomalous dimension for the BMN states first derived in \cite{SZ} (see also \cite{Beisert:2005tm} for a derivation based on integrability of the $N=4 $ spin chain). If we had used a similar saddle point intuition for droplet states as was done in \cite{CCS} we would have obtained the same result. 
Notice that these states become identified with giant magnons in the string sigma model \cite{HM}, which are also described in the Maldacena-Hofman paper as if they end on light-like  D-branes.  This geometry also helps elucidate aspects of the  integrable spin chain model  (see \cite{Beisert:2006qh} for example). Here we see that the transition from strings ending on giant gravitons to giant magnons seems to be smooth, so the distinction between an object being made of branes versus it being just strings is smooth. Remember that in our approximations the energy of the giants is bigger than order $\sqrt N$, whereas for planar diagrams in closed string states the $R$-charge has to be smaller than $\sqrt N$. There seems to be no problem for interpolating between these large charge and small charge limits. 

Now let us interpret this result in a different context, that of emergent gauge symmetry. As discussed previously, when we have giant gravitons states there is an emergent gauge symmetry for each giant graviton, so if we have two giant gravitons we expect in general a $U(1)\times U(1)$ symmetry. If the giant gravitons are of the same type and they are on top of each other, we expect to get an enhanced $U(2)$ gauge symmetry \cite{DLP,Pol}, because in the gravity theory they are described as D-branes \cite{MST}. When we separate the D-branes the
 off-diagonal modes should be thought of as $W$ bosons for an enhanced $U(2)$ symmetry broken down to $U(1)\times U(1)$. The Higgs contribution to the mass of these bosons would be associated to a Higgs condensate proportional to $v \propto |\lambda_1-\lambda_2|$. 
 The problem of doing that directly in the orbifold setup is that there is no point of enhanced gauge symmetry, but the geometric intuition of these states as getting a mass from separating the position of two branes gives the same result, and in the particular case of orbifolds the extra mass associated to the brane separation can be interpreted as the Higgs mechanism in the covering theory before the projection. We can imagine that if the usual rules of string theory related to orbifolds are applicable in these instances of field theory, then the calculation of strings stretching between two giants in the ${\cal N}=4$ SYM will give the same answer as above. Sorting this out requires understanding the normalizations of multi-giant states in more detail than what has been presented here, but there is no in principle obstacle to believe that this is not the case.
Notice that this type of calculation can be generalized to more general orbifolds and other superconformal field theories in four dimensions. Presumably, the same type of techniques would also work for perturbative theories in lower dimensions like the ABJM model \cite{ABJM} with a little more effort (the saddle point techniques based on eigenvalues suggest that this should be possible in this case as well \cite{BTran}).  Studying these other setups is beyond the scope of the present paper.

\section{Open spring theory revisited}\label{sec:ost}

The authors in \cite{deMelloKoch:2011ci} have considered understanding the Hamiltonian of giant gravitons with a fixed number of strings attached between them and they found after a lot of work that the problem of diagonalizing the one loop anomalous dimensions for this problem could be studied in terms of a system of harmonic oscillators, where the spring constants depend on the number of strings stretched between different giant gravitons. The original description of these states was given in \cite{BBFH}, and their systematic study started in \cite{dMK:2007uu,dMK:2007uv,Bekker:2007ea}.
Their computation methods use the full power of Schur-polynomials and the restricted Schur polynomial combinatorial technology that various groups have developed over the years. These efforts have culminated in descriptions of effective Hamiltonians for these states (see \cite{ Koch:2011hb,Carlson:2011hy} and references therein for a complete history) that are described in terms of harmonic oscillator degree of freedom. Their efforts required a huge amount of combinatorial prowess in order to get to these results. A  systematic technology to deal with multi-matrix correlators has also been developed, with work starting in 
\cite{Kimura:2007wy,Brown:2007xh}. 

What I will do now is rederive the `spring field theory' results using the results in the previous sections to show the power of the collective coordinate approach that has been developed in this paper. I will assume that the rules for dealing with multiple giants in the collective coordinate reproduce the results in \eqref{eq:stringstretch} for each pair of giants and each string stretching between them. The reader should notice that adding multiple strings stretching between two giants does not seem to pose too much extra work than what we did already in this paper. This is because the generating series found in the appendix \eqref{eq:main1} suggests that adding more strings will result in a polynomial multiplying the exponential, and the exponential will dominate in the planar limit for all calculations. What we find then is that if we have a collection of $n_{ij}$ strings stretching between the giants $i,j$, then the
one loop anomalous dimensions for these states should be
\begin{equation}
A n_{ij} g_{YM}^2 |\lambda_i-\lambda_j|^2
\end{equation}
whereas their classical contribution to the dimension is $n_{ij}$.
Here $A$ is a normalization constant that is a number (factors of $\pi$ etc, but no dependence on $N$), and depends on our field theory definitions of the coupling constant. The simplest way to normalize it is to match it with the giant magnons \cite{SZ,HM} as described in the previous section. For what we need, all of this can effectively be absorbed in $g_{YM}^2$ itself. The same factor of $n_{ij}$ was shown to be there by combinatorial methods in \cite{deMelloKoch:2012ck}.

Notice that the R-charge of each giant is matched with $N-|\lambda_i|^2$. If we fix the R-charge of a state to be equal to $L$, then we have that
\begin{equation}
L = \sum_i (N- |\lambda_i|)^2
\end{equation}
and this is constant, and also equal to the energy of the giant gravitons in the absence of strings, as shown previously in the computation of the effective action for a single giant graviton in equation \eqref{eq:effaction}.  
The free motion of the giant gravitons is such that the norm $|\lambda_i-\lambda_j|$ does not change in time. This means that in the presence of strings, the masses of the strings do not change with time in the absence of back reaction. Back reaction on D-branes is suppressed by their tension, so the string modes are adiabatic and we have that $\dot n_{ij}=0$. Thus we can consider their contribution to the full Hamiltonian in an adiabatic approximation and it will be given as follows
\begin{equation}
H_{total}= \sum_i (N- |\lambda_i|)^2+ \sum_{ij} n_{ij} g_{YM}^2 |\lambda_i-\lambda_j|^2
\end{equation}
Whereas the full action is
\begin{equation}
S_{eff}= \int dt  \sum_k i \lambda_k^* \dot \lambda_k - \sum_i (N- |\lambda_i|)^2- \sum_{ij} n_{ij}( g_{YM}^2 |\lambda_i-\lambda_j|^2+1) \label{eq:effspringaction}
\end{equation}
We thus find that the full action is quadratic in the $\lambda$ collective coordinates. This is a first order formulation of harmonic oscillators, so there are as many harmonic oscillators as there are giant gravitons. For this to make sense, all the non-planar diagrams that were thrown away before need to give negligible contributions throughout the full trajectories that these oscillators describe (this is, string splitting and joining needs to be supressed for long times).
The diagonalization of the quadratic problem
leads to the same frequencies as those studied in \cite{deMelloKoch:2011ci}.  The fact that the corrections to the frequency of motion relative to $1$ (the natural motion of the D-branes in the absence of strings) are of order $g_{YM}^2$ shows that backreaction is suppressed as expected.  
 The advantage of this perspective is that the oscillator motion has a clear geometric interpretation in terms of positions of holes in the fermion droplet.
 
At strong coupling we would expect that the quadratic term we have gets replaced by a square root type formula,
\begin{equation}
H_{strings} \simeq \sum n_{ij}\sqrt {1+g_{YM}^2  |\lambda_i-\lambda_j|^2}
\end{equation}
which is not expected to lead to an integrable system when substituting the corresponding terms in \eqref{eq:effspringaction}. 

At this stage we might wonder why was the result so easy to write? There are technical reasons for this. First, there is a huge degeneracy of states with the same $L$ angular momentum. If we pick a basis where each giant graviton has a fixed quantized angular momentum, we get the representation of the Hamiltonian in terms of Young Tableaux and all the technology developed leading to \cite{deMelloKoch:2011ci} is necessary to deal with the operator mixing. In our case, by introducing the parameters $\lambda$ we have chosen a basis of states that is not diagonal in R-charge for each giant anymore, but we have traded that for the presence of zero modes in the sense that we have phases for the $\lambda_i$ that do not change the energy of a state, but that change its location. The relative angles matter. What we see is that the one loop correction we computed is local in the zero modes: giants with  different phase angles are orthogonal to each other, even if we have finite excitations on them, because of the exponential suppression of the overlaps that is standard for coherent states in the harmonic oscillator. Moreover the string Hamiltonian breaks the degeneracy between the zero modes, thus the effective eigenstates of the perturbation Hamiltonian for the zero modes would correspond to localizing at fixed values of  the zero modes themselves. If the basis has already done that, the effective Hamiltonian is diagonal and mixing of states does not have to be computed. This is why the introduction of these collective coordinate results in a huge simplification at the end of the day. We can  quantize these collective coordinates again if we want to find the fine spectrum of the Hamiltonian, but in the case of harmonic oscillators this is essentially trivial.

What we will do instead is look at special configurations. Consider first the case of two giant graviton operators. The equation of motion then give us that 
\begin{eqnarray}
\dot \lambda_1 =  i \lambda_1 -i n_{12} g_{YM}^2( \lambda_1-\lambda_2)\\
\dot \lambda_2 =  i \lambda_2 -i n_{12} g_{YM}^2( \lambda_2-\lambda_1)
\end{eqnarray}
It is easy to see that the center of mass motion for the coordinate $\lambda_1+\lambda_2$ does not change. The other mode is
\begin{equation}
(\dot \lambda_1-\dot \lambda_2) = i (\lambda_1-\lambda_2) - 2 i n_{12} g_{YM}^2( \lambda_1-\lambda_2)
\end{equation}
which has a frequency equal to $1- 2 n_{12} g_{YM}^2$. Thus the center of mass will circle the origin at the same angular rate as if there were no strings. Whereas the relative position of the branes will circle around the center of mass at a slightly slower rate. This means that if one of the giants can graze the edge of the droplet, it will move at an angular frequency that is lower than the edge excitations. This is, adding strings is like giving a mass to the D-branes in the plane wave limit and they therefore travel slower than the gravitons (the giants without strings), which travel at the speed of light. It would also be interesting to see such a frequency shift in a direct calculation of supergravity with giant gravitons and short strings suspended between them. 

We can also consider what happens when one such giant graviton reaches the edge of the droplet. Because the approximations we have done  to get to the effective action break down, the giant graviton can not penetrate to the other side as a hole, and it must pass through a region where it has dissolved into strings. Thus in that limit the configuration connects to a collection of  strings attached to one fewer giant graviton, and these strings are not simple impurities ( a single $Y$) as we have described here, but require solving the associated spin chain model ending on branes. It is unclear whether such object can reconstitute itself as a D-brane in a short time, as the entropy of the configuration of strings is larger.

 \section{Outlook}\label{sec:outlook}

This paper has produced a set of calculations for giant gravitons by introducing a collective coordinate and making use of the collective coordinate states so that a lot of the difficulties in dealing with giant graviton states end up being dealt with by a simple generating series. This paper is in some sense a rewriting of a lot of previous works of the author \cite{Bstudy,Btoy,BBFH} with a prescription to do calculations and it also shows that the na\"\i ve localization techniques of \cite{BCV} can be made rigorous in the case of giant gravitons with strings suspended between them. 
Notably, the fermion droplet picture of \cite{Btoy} comes alive and it describes exactly the geometry that the giant gravitons are allowed to explore. In particular a first order formulation of the action of the giant graviton state was derived by using a Berry phase calculation and the known energies of the states. In this paper I was also able to give a clean geometric interpretation to the `spring field theory' calculations \cite{deMelloKoch:2011ci} so that the springs describe the motion of the giant gravitons in the fermion droplet picture. In this setup one treats the $Y$ impurities as adiabatic modes, and one can analyze the back reaction of the D-branes (giant gravitons) in a controlled setting.

It is obvious that the techniques presented here can be used in other setups. For example, one can study marginal deformations of ${\cal N}=4$ SYM, which are known to give rise to interesting deformations of the $AdS_5\times S^5$ geometry and to interesting orbifold spaces  \cite{BJL}, where there is some non commutativity in the $SO(6)$ directions. The gravity solutions were found in \cite{Lunin:2005jy}. They also lead to interesting deformations of the ${\cal N}=4 $ spin chain \cite{Berenstein:2004ys}. Understanding the geometry of these configurations with probe D-branes would certainly be interesting in the field theory setup.
Other interesting setups are those of orbifolds of $N=4 $ SYM away from the orbifold point as in \cite{Gadde:2010ku}. The techniques presented here should be readily applicable to those setups and one should be able to observe the dispersion relation of the twisted magnons arising simply from a calculation along the lines of this paper. In a general sense, one should be able to start detecting the RR field strengths present in the gravity theory as well.

Another obvious application is to revisit the problem of open spin chains attached to giant gravitons \cite{Berenstein:2005vf,Berenstein:2005fa,Berenstein:2006qk}. Now that the giant gravitons are parametrized by a coordinate $\lambda$, the boundary condition that these induce on the spin chain of $N=4$ SYM is parametrized by the complex variable $\lambda$, rather than by the number of boxes in a Young tableaux . In these works it was found that the boundary condition could change the number of sites on a spin chain if interpreted in terms of the usual $SU(2)$ spin degrees of freedom, but that it could also be thought of as a bosonic lattice model with hopping between sites with non-diagonal boundary conditions.
 It would be nice to analyze the ground states of these spin chains 
and to see if there is a reasonable spacetime interpretation of the results obtained that way. The end result should look very similar to \cite{Hatsuda:2006ty}, which used the saddle point techniques of \cite{BCV} to derive their results. One could hope that in this way one would build the field theory degrees of freedom for the Higgs phase for the field theory of two giant gravitons separated from each other, with the correct dispersion relation for a relativistic massive field on an $S^3$. This is also related to the bound state problem of giant magnons \cite{Dorey:2006dq}.

For many of these computations one really needs to work with two giant graviton states of the same gauge group, of the form $\det(Z-\lambda_1) \det(Z-\lambda_2)$. It is expected that the generating series technique that worked so well for one giant graviton in this work can be extended to the case of two (or many such giant gravitons) without too much extra work. Also, it would be nice to understand better how the nonabelian enhancement of symmetry from $U(1)^2$ to $U(2)$ happens dynamically in the interactions of strings attached to the giant gravitons. I believe that the expressions should be simple sums of Schur polynomials \cite{CJR} for which an approximation similar to the ones found in this paper would apply. This probably easiest to do with fermion wave functions in the holomorphic representation, which are equivalent to the Young Tableaux formalism.

Clearly, it is also interesting to study the splitting and joining of strings attached to giants, as well as to carry out the calculations of the anomalous dimensions to higher orders in perturbation theory. Presumably one will find that the geometry of the strings attached to giant gravitons can become sensitive to the full $S^5$ geometry. Naturally, one should study how this works for orientifolds, where apart from determinants and traces one also has Pfaffian operators \cite{Witten:1998xy}. Some recent progress in the `Schur basis' description for this setup can be found in \cite{Caputa:2013hr}. 
I also think that it is possible to improve on the description of magnons for more general droplet configurations \cite{CCS} so that one does not rely on a saddle point calculation to get the end result. 

 It would also be nice to find a collective coordinate for giants growing into $AdS_5$ that has similar properties to the collective coordinate $\lambda$. This is, that it is restricted to be larger than the radius of the fermion droplet. A na\"\i ve such coordinate is the eigenvalues of $Z$ themselves, but what we want is a parameter that gives a reasonable generating series similar to those found in section \ref{sec:colcoor} and those found in the appendix, such that a nice approximation to the appropriate generating series exists for $|\lambda|>\sqrt N$ and which gives rise to a first order realization of a  harmonic oscillator.

Finally, it should be noted that it is not hard to take a plane wave limit of the giant graviton configurations, even with strings stretching between them. Doing this carefully might shed some light on the 
`tiny graviton matrix theory' conjectures \cite{SheikhJabbari:2004ik}.  

\acknowledgments

Work supported in part by DOE under grant DE-FG02-91ER40618. 

\appendix

\section{Notation and normalization of correlators}

For completeness, this appendix defines the notation and some basic correlators. This follows the notation introduced in
\cite{Bstudy}. The first important identity is the following equation which is a definition of the determinant operator for $M$. Assume that $M$ is  an $n\times n$ marix, the the determinant is given by
\begin{equation}
\det(M) = \frac 1 {n!} \epsilon_{i_1, \dots ,i_n} \epsilon^{j_1, \dots j_n} M^{i_1}_{j_1} \dots M^{i_n}_{j_n}
\end{equation}
The combination 
\begin{equation}
\epsilon_{i_1, \dots ,i_n} \epsilon^{j_1, \dots j_n}
\end{equation}
is an invariant tensor for the $GL(n,\BC)$ group. We will notate this combination as the $\epsilon\epsilon$ tensor. Given any matrix $M^{i_1}_{j_1}$, we think of it as a tensor of $GL(n,\BC)$ with one upper and one lower index. We can contract either one or both pairs of the indices of $M$ with those of the $\epsilon\epsilon$ tensor. If we contract both tensor indices, we want to use a notation that does this automatically without writing indices in the first place. Here, it helps that $\epsilon_{i_1, \dots ,i_n}$ and $\epsilon^{j_1, \dots j_n}$ are completely antisymmetric in their indices. Thus picking any upper index in $\epsilon\epsilon$ can be permuted to the first upper index. Similarly with a lower index. This means that there is essentially only one way to do this contraction (up to a sign from the permutation) and this is given by
\begin{equation}
\epsilon_{i_1, \dots ,i_n} \epsilon^{j_1, \dots j_n} M^{i_1}_{j_1}
\end{equation}
We will use the following notation for this combination
\begin{equation}
\epsilon\epsilon(M) ^{j_2, \dots j_n}_{i_2, \dots i_n}= \epsilon_{i_1, \dots ,i_n} \epsilon^{j_1, \dots j_n} M^{i_1}_{j_1}
\end{equation}
Similarly, if we have multiple matrices $M^1, M^2, \dots, M^k$, we can define the following multilinear tensor
\begin{equation}
\epsilon\epsilon(M^1, \dots, M^k)^{j_{k+1}, \dots, j_n}_{i_{k+1}, \dots, i_n}=\epsilon_{i_1, \dots ,i_n} \epsilon^{j_1, \dots j_n} (M^1)^{i_1}_{j_1}\dots (M^k)^{i_k}_{j_k}
\end{equation}
when $k=n$ there are no free indices and we get a number.  This is denoted by
\begin{equation}
\epsilon\epsilon( M^1, \dots , M^n)
\end{equation}
we can use this same notation if we only saturate $s$ indices inside $\epsilon\epsilon$, so long as it is understood that there are $n-s$ indices floating. If we want to be pedantic we can call these objects $\epsilon\epsilon_{n-s}(M^1, \dots, M^s)$. 

We now get the obvious relation
\begin{equation}
\det(M) = \frac 1 {n!} \epsilon\epsilon(M, \dots M)
\end{equation}
The $\epsilon\epsilon$ tensor is obviously linear in the entries, so that
\begin{equation}
\epsilon\epsilon(  \alpha M^1 +\beta M^2, \dots)= \alpha \epsilon\epsilon(M^1, \dots)+\beta\epsilon\epsilon(M^2, \dots)
\end{equation}
It is also invariant under permutations of the $M^i$. The sign that we obtain from permitting upper indices gets canceled because we do the same permutation in the lower indices. 
From this it follows that if all entries are linear combinations of $M^1, M^2$ then
\begin{equation}
\epsilon\epsilon( \alpha M^1 +\beta M^2, \dots,  \alpha M^1 +\beta M^2)
=\sum_{k=0}^n {n \choose k}  \alpha^k \beta^{n-k} \epsilon\epsilon( \underbrace{M^1\dots M^1}_{k \hbox{ times}} ,\underbrace{ M^2, \dots, M^2}_{n-k \hbox{ times}}) 
\end{equation}
Similar binomial identities apply if we have only $s$ indices saturated by the combination $ \alpha M^1 +\beta M^2$, and the other $n-s$ indices are left floating. These are given by
\begin{equation}
\epsilon\epsilon_{n-s}( \alpha M^1 +\beta M^2, \dots,  \alpha M^1 +\beta M^2)
=\sum_{k=0}^s {s \choose k}  \alpha^k \beta^{s-k} \epsilon\epsilon_{n-s}( \underbrace{M^1\dots M^1}_{s \hbox{ times}} ,\underbrace{ M^2, \dots, M^2}_{n-k-s \hbox{ times}}) 
\end{equation}
The main example in this paper include the sub determinant expressions given by
\begin{equation}
\det\!_k(Z) = \frac 1 {N!} { N \choose k} \epsilon\epsilon( \underbrace{Z, \dots Z}_{k \hbox{ times}}, 1, \dots 1)
\end{equation}
where the matrix $1$ has as components the Kronecker $\delta^\mu_\nu$ symbol. 
These expressions are written for $N\times N$ matrices, and the other simple example that is relevant is given by
\begin{equation}
\epsilon\epsilon( \underbrace{Z, \dots Z}_{k \hbox{ times}}, Y,1,  \dots 1)
\end{equation}

Regarding correlators of these objects, we take one such operator $Z$ and and its complex conjugate, labeled $\bar Z$. The main correlators to compute are given by
\begin{equation}
\langle \bar Z^{i}_{j} Z^{l}_m\rangle = \delta^i_m \delta^l_m 
\end{equation}
and a similar expression for $Z,Y$ matrices.
A straightforward evaluation performed in \cite{Bstudy} shows that
\begin{equation}
\langle \epsilon\epsilon( \underbrace{\bar Z, \dots \bar Z}_{N-k \hbox{ times}}, 1, \dots 1) \epsilon\epsilon( \underbrace{Z, \dots Z}_{N-k \hbox{ times}}, 1, \dots 1)=  (N-k)! ( N! (N-k)!k!)
\end{equation}
The first factor of $(N-k)!$ counts all the possible different contractions between the $Z$ and the $\bar Z$. Each of these gives the same result. The rest is a numerical identity between contractions of four $\epsilon$ tensors. 
Similarly, we  have that 
\begin{equation}
\langle \epsilon\epsilon( \underbrace{\bar Z, \dots ,\bar Z}_{N-k -1\hbox{ times}},\bar Y, 1, \dots 1) \epsilon\epsilon( \underbrace{Z, \dots ,Z}_{N-k-1 \hbox{ times}}, Y, 1, \dots 1)=  (N-k-1)! ( N! (N-k)!k!)
\end{equation}
as there are only $(N-k-1)!$ contractions between the $Z$, and only one contraction between the $Y$, but we get the same identity with four epsilon tensors as before. Similarly
\begin{equation}
\langle \epsilon\epsilon(\underbrace{\bar Z, \dots ,\bar Z}_{N-k -s\hbox{ times}},\underbrace{\bar Y, \dots,\bar Y}_{s \hbox{ times}}, 1, \dots 1) \epsilon\epsilon(\underbrace{Z, \dots ,Z}_{N-k-s \hbox{ times}}, \underbrace{\bar Y, \dots,\bar Y}_{s \hbox{ times}}, 1, \dots 1)=  (N-k-s)! s! ( N! (N-k)!k!)
\end{equation}

An important object for us is the formal sum
\begin{equation}
\det(Z-\lambda)= \frac 1{N!} \sum_{k=0}^N { N\choose k} (-\lambda)^k \epsilon\epsilon( \underbrace{Z, \dots Z}_{k \hbox{ times}}, 1, \dots 1)
\end{equation}
The correlators of these objects is
\begin{equation}
\langle \det(\bar Z-\tilde \lambda^*) \det( Z-\lambda)\rangle= N! \sum_{k=0}^N \frac{(\tilde\lambda^*\lambda)^k}{k!}\simeq N! \exp(\tilde \lambda^*\lambda)
\end{equation}
which gives a truncated exponential.
Moreover, we can take derivatives with respect to $\lambda$. For example
\begin{equation}
\partial_\lambda \det(Z-\lambda) = \det(Z-\lambda) \Tr \frac{-1}{Z-\lambda} 
\end{equation}
and
\begin{equation}
(\partial_\lambda)^2 \det(Z-\lambda) = \det(Z-\lambda) \left[\left(\Tr \frac{1}{Z-\lambda}\right)^2 - \Tr\frac {1}{(Z-\lambda)^2}\right]
\end{equation}
and we get the correlations
\begin{equation}
\langle \det(\bar Z-\tilde \lambda^*)\det(Z-\lambda) \Tr \frac{-1}{Z-\lambda} \rangle \simeq \tilde \lambda^* \exp(\tilde \lambda^*\lambda)
\end{equation}
etc.

Similarly, consider the sum given by
\begin{equation}
\det(Z-\lambda)\Tr\left( Y \frac 1 {Z-\lambda}\right)= \frac 1{(N-1)!}\epsilon\epsilon( Z-\lambda, \dots, Z-\lambda, Y)
\end{equation}
We can expand this expression using our binomial expression as follows
\begin{equation}
\frac 1{(N-1)!}\epsilon\epsilon( Z-\lambda, \dots, Z-\lambda, Y)= \frac 1{(N-1)!}\sum {N-1 \choose k} (-\lambda)^k \epsilon\epsilon( \underbrace{Z, \dots Z}_{N-k-1 \hbox{ times}}, Y,1,  \dots 1)
\end{equation}
Sandwiching this sum in correlators gives us
\begin{eqnarray}
&\langle\det(\bar Z-\tilde\lambda^*)\Tr\left(\bar Y \frac 1 {Z-\tilde\lambda^*}\right)\det(Z-\lambda)\Tr\left( Y \frac 1 {Z-\lambda}\right)\rangle &\\
= &\sum \frac 1{((N-1)!)^2}(\tilde \lambda^* \tilde \lambda) {N-1 \choose k}^2 (N-k-1)!N!(N-k)! k!&\\
=&\sum N! (\tilde\lambda^*\lambda)^k \frac {N-k}{k!}= N! ( N-\lambda\partial_\lambda) \sum \frac{(\tilde \lambda^*\lambda)^k}{k!}&
\\
\simeq & N! (N-\tilde\lambda^*\lambda)\exp(\tilde \lambda^* \lambda)& \label{eq:main1}
\end{eqnarray}
Similarly, we can take derivatives with respect to $\lambda$ to obtain
\begin{equation}
\langle\det(\bar Z-\tilde\lambda^*)\Tr\left(\bar Y \frac 1 {Z-\tilde\lambda^*}\right)\partial_\lambda\left(\det(Z-\lambda)\Tr\left( Y \frac 1 {Z-\lambda}\right)\right)\rangle \simeq N! \tilde \lambda^* (N-\tilde\lambda^*\lambda)\exp(\tilde \lambda^* \lambda)
\end{equation}
where we have taken the leading term at large $\sqrt{N} >|\lambda|, |\tilde \lambda|>>1$ that is appropriate for the planar limit. The natural size for $\lambda$ is of order $\sqrt N$ in order for the approximation to the exponential to be valid. 
 
This derivative is given by
\begin{equation}
\partial_\lambda \left(\det(Z-\lambda)\Tr\left( Y \frac 1 {Z-\lambda}\right)\right)= \det(Z-\lambda) \left[\Tr\left(  Y \frac 1{(Z-\lambda)^2}\right) - \Tr\left( Y \frac 1 {Z-\lambda}\right) \Tr\frac 1{Z-\lambda}\right] \label{eq:main2}
\end{equation}

\end{document}